\documentclass[conference]{IEEEtran}
\IEEEoverridecommandlockouts

\usepackage{cite}
\usepackage{amsmath,amssymb,amsfonts}
\usepackage{graphicx}
\usepackage{textcomp}
\usepackage{xcolor}
\usepackage{array}
\usepackage{booktabs}
\usepackage{tabularx}
\usepackage{url}
\newcolumntype{Y}{>{\raggedright\arraybackslash}X}
\newcommand{\turn}[1]{\textcircled{\raisebox{-0.35ex}{\scriptsize #1}}}
\newcommand{\speaker}[1]{\textsc{#1}:}
\newcommand{\usertext}[1]{\textit{#1}}

\begin{document}

\title{StrokeGuard: A Multi-Agent Guided System for Prehospital Stroke Assessment}

\author{
\IEEEauthorblockN{Wentao Yang}
\IEEEauthorblockA{\textit{Institute of Medical Robotics}\\
\textit{School of Biomedical Engineering}\\
\textit{Shanghai Jiao Tong University}\\
Shanghai, China\\
tao615@sjtu.edu.cn}
\and
\IEEEauthorblockN{Zhenye Xu}
\IEEEauthorblockA{\textit{Institute of Medical Robotics}\\
\textit{School of Biomedical Engineering}\\
\textit{Shanghai Jiao Tong University}\\
Shanghai, China\\
lingyy1109@gmail.com}
\and
\IEEEauthorblockN{Ruoyi Li}
\IEEEauthorblockA{\textit{Institute of Medical Robotics}\\
\textit{School of Biomedical Engineering}\\
\textit{Shanghai Jiao Tong University}\\
Shanghai, China\\
rowenalee@sjtu.edu.cn}
\and
\IEEEauthorblockN{Musen Zhang}
\IEEEauthorblockA{\textit{Institute of Medical Robotics}\\
\textit{School of Biomedical Engineering}\\
\textit{Shanghai Jiao Tong University}\\
Shanghai, China\\
musen\_zhang@sjtu.edu.cn}
\and
\IEEEauthorblockN{Yao Guo}
\IEEEauthorblockA{\textit{Institute of Medical Robotics}\\
\textit{School of Biomedical Engineering}\\
\textit{Shanghai Jiao Tong University}\\
Shanghai, China\\
yao.guo@sjtu.edu.cn}
}

\maketitle

\begin{abstract}
Prehospital stroke assessment aims to accurately identify stroke symptoms and make rapid decisions through standardized procedures within an extremely narrow time window, thereby saving valuable time for subsequent treatment. In clinical practice, FAST-based scales are widely used for prehospital stroke assessment by issuing instructions that guide subjects to perform specific actions to screen facial, arm, and speech functions. However, in home and community settings, non-clinical users often encounter challenges such as inaccurate descriptions, incomplete symptom observation, and difficult operational procedures, which may lead to inaccurate or biased assessment results. To address these challenges, this paper presents StrokeGuard—a multi-agent guided system designed for prehospital stroke assessment that makes mobile FAST screening more standardized and executable. Specifically, to overcome the limitations of traditional single-agent systems in terms of procedural fault tolerance and user guidance capability, StrokeGuard adopts a dual-channel agent mechanism that separates formal assessment (i.e.., facial palsy, arm weakness, speech impairment) from procedural support (e.g., step prompts, error correction, and real-time feedback). It guides the assessment process through multi-agent collaboration, dual-channel interaction, state-machine control, and stage-local fallback recovery mechanisms. Stage-specific scoring is delegated to constrained pretrained video assessment modules, while evidence source records are integrated with structured report generation. The user evaluation uses MATES-9, an exploratory scale for measuring user experience in multistep AI-guided tasks. In a simulated prehospital scenario, StrokeGuard improves the MATES-9 total score over a paper FAST-style form by 10.83 points, corresponding to a 23.8\% relative increase.
\end{abstract}


\section{Introduction}

Stroke is an acute neurological event caused by a sudden interruption of cerebral blood flow or rupture of a cerebral vessel. Because brain injury progresses rapidly, prehospital stroke assessment must identify symptoms accurately and support rapid decisions through standardized procedures within an extremely narrow time window. Prehospital screening tools such as the Cincinnati Prehospital Stroke Scale (CPSS) and the Face Arm Speech Test (FAST) reduce stroke recognition to observable facial, arm, and speech signs \cite{ref1,ref2,ref3}. In clinical practice, FAST-based scales guide subjects to perform specific actions so that facial, arm, and speech functions can be screened in a structured manner.

The apparent simplicity of these tools does not directly translate into reliable execution by low-expertise users. In home and community settings, non-clinical users often face inaccurate descriptions, incomplete symptom observation, and difficult operational procedures. Prehospital studies report clear limits in stroke recognition accuracy even when dispatchers and emergency medical personnel participate \cite{ref8,ref10}. Barriers to timely activation of emergency medical services also include difficulty interpreting symptoms, insufficient judgment of severity, and uncertainty about appropriate actions \cite{ref11}. When FAST-style screening moves from trained personnel to ordinary caregivers, a system must support more than recognition of the three FAST categories. It must also provide procedural guidance, support multimodal evidence collection under time pressure, and preserve the current assessment step after capture failure or workflow interruption.

\begin{figure}[!t]
\centering
\includegraphics[width=\linewidth]{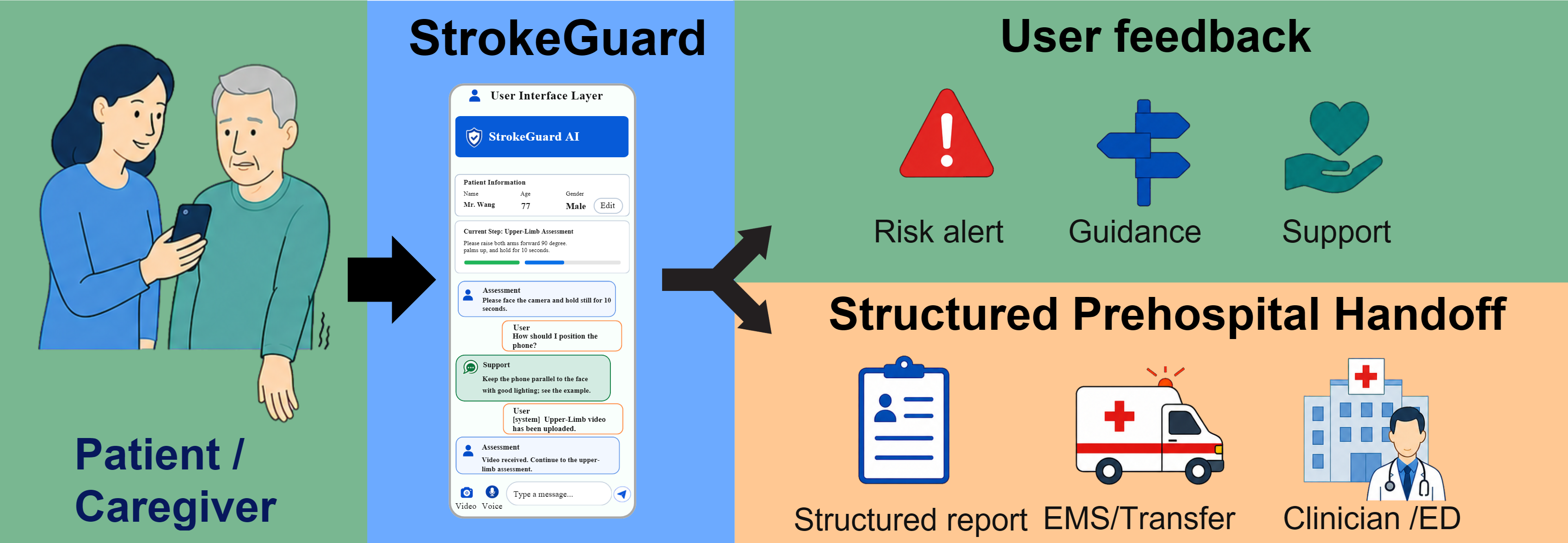}
\caption{High-level overview of StrokeGuard in a simulated prehospital assessment workflow.}
\label{fig:intro-overview}
\end{figure}

Recent multimodal stroke systems have advanced perception and classification. DeepStroke uses multimodal adversarial learning to support emergency-room stroke screening \cite{ref4}. M3 Stroke studies mobile multimodal AI for mild-to-moderate acute stroke triage \cite{ref5}. Digital FAST explores a multimodal framework aligned with the FAST paradigm \cite{ref6}. VOICE demonstrates the feasibility of a voice AI agent for simulated prehospital stroke assessment \cite{ref7}. These studies show that AI can support early triage, but they do not fully address the interaction and handoff problems that arise before a model receives usable input. An assessment model alone cannot guide users through recording failures or preserve workflow order. A general conversational agent can provide flexible dialogue, but it does not by itself guarantee controlled scoring, adequate multimodal input quality, or an auditable handoff for clinicians.

This execution gap is also distinct from the problems addressed by specialized prehospital solutions. Mobile stroke units show that richer prehospital data collection can improve triage and reduce door-to-needle time, but these systems rely on trained onboard personnel rather than ordinary caregivers at the scene \cite{ref12}. Dispatcher recognition algorithms and structured call-taking protocols have been used to improve telephone-based stroke screening, but they target trained dispatchers rather than direct interactive guidance for untrained on-site users \cite{ref13}.

To address these challenges, this paper presents StrokeGuard, a multi-agent guided system designed for prehospital stroke assessment that makes mobile FAST screening more standardized and executable. To overcome the limitations of traditional single-agent systems in procedural fault tolerance and user guidance capability, StrokeGuard adopts a dual-channel agent mechanism that separates formal assessment of facial palsy, arm weakness, and speech impairment from procedural support, including step prompts, error correction, and real-time feedback. It guides the assessment process through multi-agent collaboration, dual-channel interaction, state-machine control, and stage-local fallback recovery. Stage-specific scoring is delegated to constrained pretrained video assessment modules, while evidence-source records are integrated with structured report generation.

This paper makes three contributions:
\begin{itemize}
\item A prehospital stroke assessment agent system. StrokeGuard reframes FAST-style screening for low-expertise users as a multi-agent assessment execution problem and places formal assessment, procedural support, stage progression, local fallback, and source-preserving handoff within explicit responsibility boundaries.
\item A controlled execution architecture for assessment agents. The architecture separates agent interaction, deterministic orchestration, and dedicated scoring modules, allowing stage-specific modules to be replaced by different stroke recognition algorithms and supporting transfer to other clinical screening tasks with fixed procedures or multiple assessment items.
\item An evaluation design for controlled assessment agents. The paper uses mechanism-ablation case studies to examine the roles of dual-channel constraints, evidence-aware orchestration, and stage-local recovery, and proposes MATES-9 as an exploratory tool for evaluating the experience of multistep AI-guided tasks.
\end{itemize}

\section{Related Work}

\subsection{Multimodal AI Algorithms for Stroke Assessment}

Prior multimodal stroke-screening studies show that facial video, speech, and limb motion can support automatic identification of stroke-related signs. DeepStroke, M3 Stroke, and Digital FAST validate automatic stroke screening from face and speech cues, audiovisual temporal features, and FAST-component modeling, respectively, and report strong sensitivity, accuracy, or F1 scores \cite{ref4,ref5,ref6}. Sensor-based work has also explored stroke-related functional assessment, including balance-function assessment with a single ear-worn IMU \cite{ref42}.

Taken together, this line of work establishes facial dynamics, arm motion, speech signals, and wearable sensing as useful evidence streams for stroke-related assessment. However, existing studies mainly examine feature extraction, modality fusion, and classification or functional-estimation performance, often assuming that input of acceptable quality is already available. StrokeGuard does not retrain a stroke classifier. It wraps pretrained Face, Arm, and Speech assessment algorithms as constrained modules and addresses input availability through staged capture guidance, video-first processing, and single-stage text fallback, while limiting the effect of capture failure on the full assessment process.

\subsection{Clinical Healthcare Agents}

Clinical healthcare agents target specific clinical or health tasks by combining language models with context maintenance, planning, tool use, memory, and feedback \cite{ref16,ref17}. In medicine, these agents have been used for diagnostic assistance, clinical decision support, documentation, patient communication, and workflow management. Their actions, however, must remain constrained by clinical authority, human supervision, safety rules, and evidence-tracing mechanisms \cite{ref16,ref17}. Medical conversational agents also support health education, symptom inquiry, and person-centered care through natural language \cite{ref14,ref15}, with evaluation concerns extending to factual reliability, clinical safety, task completion, and human-AI collaboration in real contexts \cite{ref14,ref15,ref16,ref17}. When clinicians review agent outputs, structured handoff and trust calibration are also important: structured handoff tools reduce omissions of key information \cite{ref39}, while adoption of clinical AI depends on task-relevant explanation, context, and uncertainty \cite{ref40}. A prehospital agent system should therefore preserve input sources, processing states, and result-formation processes in addition to conclusions.

\subsection{Multi-Agent Systems}

LLM-based multi-agent systems provide an engineering paradigm for role division and collaborative execution. MetaGPT and ChatDev show that the value of multi-agent systems lies not only in the number of models but also in the organization of tasks, roles, communication paths, and intermediate artifacts \cite{ref35,ref36}. For prehospital stroke assessment, this view suggests that input routing, formal assessment guidance, and procedural support can be assigned to different roles while clinical authority, safety rules, and evidence-tracing constraints remain explicit \cite{ref16,ref17}.

Role division alone is insufficient for reliable medical assessment. A system must specify which inputs can change assessment state, how state changes are recorded, which stage a media result belongs to, and whether a local failure affects completed stages. Workflow architecture provides related design cues: CQRS, Event Sourcing, Saga, and Bulkhead respectively motivate write boundaries, event traces, multistep compensation, and fault isolation \cite{ref28,ref29,ref30,ref31}. Task-oriented dialogue research also shows that natural-language systems need explicit task state and a separation between general dialogue behavior and domain task control \cite{ref18,ref32,ref33,ref34}. These issues are sharper in prehospital stroke assessment, where caregivers may face symptom interpretation, action decisions, nonprofessional operation, and emergency-response difficulties \cite{ref8,ref10,ref11}, while medical conversational agents must also satisfy contextual adaptation, user interaction, and safety requirements \cite{ref14,ref15}. StrokeGuard therefore brings multi-agent role division, workflow state boundaries, and task-oriented dialogue control into prehospital stroke assessment through dual-channel agent interaction, evidence-aware orchestration, stage-local recovery, and source-traceable handoff.

\subsection{Human-Computer Interaction Evaluation in Medical Systems}

Studies of clinical decision support and prehospital medical technology commonly evaluate task outcomes together with efficiency, cognitive burden, and usability \cite{ref19,ref20,ref21}. Existing scales provide useful but incomplete construct sources: SUS, NASA-TLX, MAUQ, and Health-ITUES cover usability, workload, mobile-health experience, and health-information-technology experience \cite{ref22,ref23,ref24,ref25}, while CUQ and BUS cover chatbot and conversational-interface usability \cite{ref26,ref27}. MATES-9 is designed as an exploratory tool for multistep AI-guided tasks by adapting these constructs to measure task guidance and step awareness, information clarity, support continuity, perceived workload, and action-decision support.

\section{StrokeGuard System Design}

\subsection{Overall Architecture}

\begin{figure*}[t]
\centering
\includegraphics[width=0.94\textwidth]{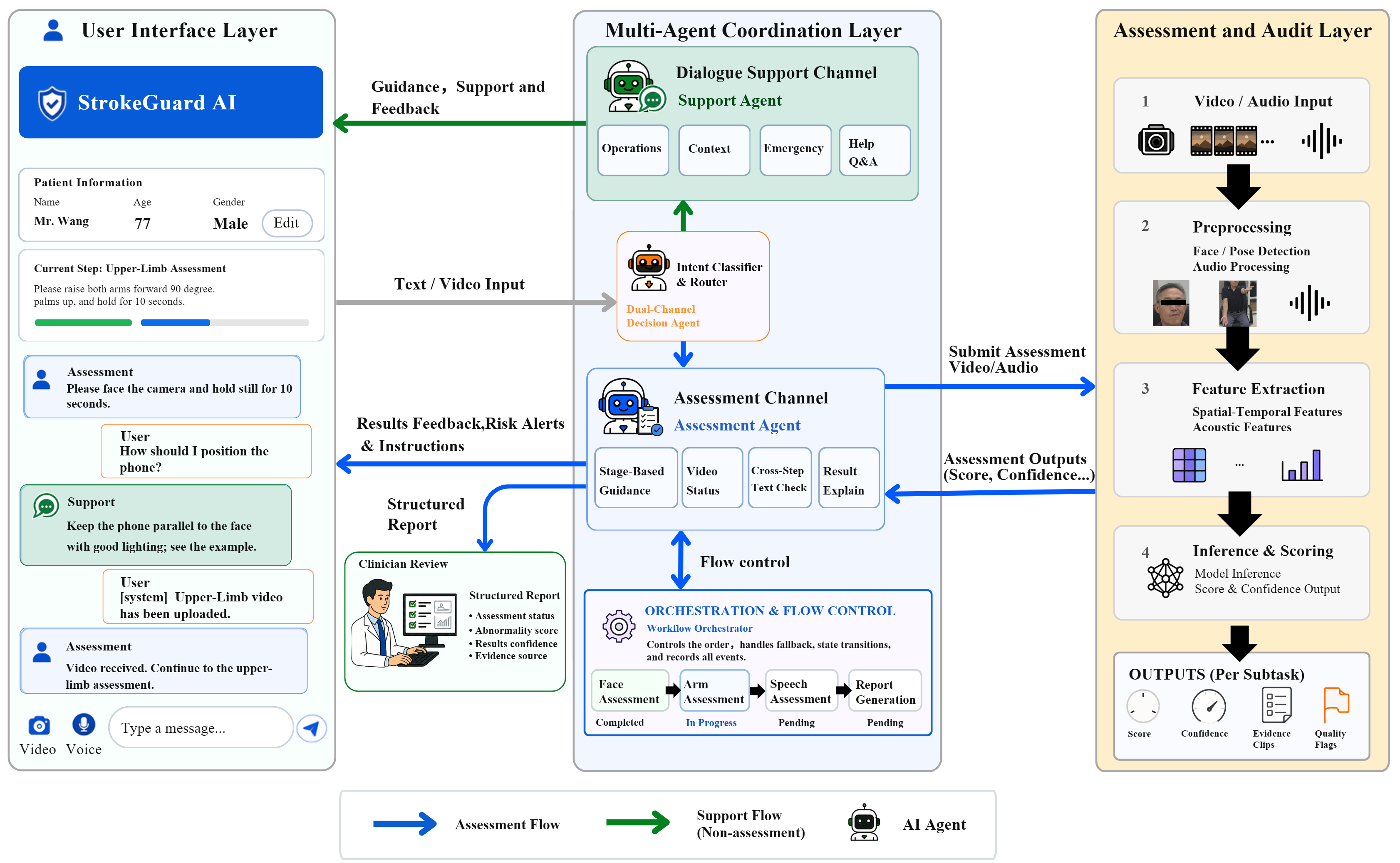}
\caption{System architecture of StrokeGuard. User inputs are routed to assessment or support channels, deterministic orchestration controls stage progression and fallback, and stage-specific assessment modules provide source-preserving outputs for clinician review.}
\label{fig:architecture}
\end{figure*}

StrokeGuard combines natural-language agents with deterministic flow control in the layered architecture shown in Fig.~\ref{fig:architecture}. The User Interface Layer presents the current stage, capture action, and supportive communication to the caregiver. The Multi-Agent Coordination Layer routes input and handles natural-language interaction while coordinating deterministic assessment-state control. The Assessment and Audit Layer calls dedicated Face, Arm, and Speech modules and stores evidence sources and processing traces.

\subsection{Dual-Channel Interaction and Assessment-State Separation}

\begin{table}[t]
\caption{Agent roles and outputs in StrokeGuard architecture}
\label{tab:roles}
\centering
\footnotesize
\setlength{\tabcolsep}{4pt}
\renewcommand{\arraystretch}{1.08}
\begin{tabular}{|p{0.33\linewidth}|p{0.57\linewidth}|}
\hline
Agent & Roles and output \\
\hline
Dual-channel decision agent & Routes input to assessment or support; outputs channel label, target role, and routing record. \\
\hline
Assessment agent & Guides current-stage capture; outputs media submission, text fallback request, completion, waiting, or failure event. \\
\hline
Support agent & Answers operational, explanatory, reassuring, and emergency-response questions; outputs support reply without state change. \\
\hline
\end{tabular}
\end{table}

The dual-channel mechanism addresses confusion that can arise when dialogue state and clinical state share the same control path. In a single-channel design, a statement such as the left arm of the patient is dropping and a question such as how should I hold the phone can enter the same conversational context, and the same model may generate both the reply and the next action. The former may constitute assessment evidence, whereas the latter should only help the user continue capture. Task-oriented dialogue research usually constrains later actions through explicit state tracking and task control \cite{ref18,ref32,ref33,ref34}. Prehospital assessment also needs a clear distinction between inputs that can affect a clinical stage and inputs that cannot. Without this distinction, a shared control path may cause repeated prompts, stage jumps, evidence misclassification, or unclear evidence sources.

StrokeGuard separates these two types of input through the dual-channel decision agent and sends them to different agent channels. The assessment-channel agent produces formal events that can be verified by the orchestrator, including media submission, text fallback completion, processing failure, and current-stage completion. The support-channel agent reads the same session context to keep replies relevant, but it has no permission to write assessment evidence or advance the assessment stage. It produces only natural-language replies and no state-transition events. Users can therefore ask support questions at any time and receive answers without disrupting assessment-stage progression.

This mechanism can be expressed as a routing function with permission constraints. Let the user input be $x_t$, the session context be $c_t$, and the clinical assessment state be $s_t$. The dual-channel decision agent outputs the routing label
\begin{equation}
r_t=R(x_t,c_t), \qquad r_t\in\{\mathrm{assessment},\mathrm{support}\}.
\end{equation}
The two channels have different permission sets:
\begin{equation}
\begin{aligned}
\mathcal{P}(\mathrm{assessment})&=\{\mathrm{create\_event},\mathrm{write\_evidence},\\
&\quad \mathrm{request\_transition}\},\\
\mathcal{P}(\mathrm{support})&=\{\mathrm{read\_context},\mathrm{generate\_reply}\}.
\end{aligned}
\end{equation}
State update occurs only when the assessment event $e_t$ passes orchestration validation:
\begin{equation}
s_{t+1}=
\begin{cases}
T(s_t,e_t), & r_t=\mathrm{assessment}\land V(e_t,s_t)=1,\\
s_t, & r_t=\mathrm{support}\lor V(e_t,s_t)=0.
\end{cases}
\end{equation}
Here, $V(e_t,s_t)$ denotes checks for stage consistency, completion conditions, and evidence conflicts. The equation captures the core constraint of the dual-channel design: the support channel can generate replies but cannot change clinical state.

The dual-channel decision agent sits between user input and the two task channels. It converts natural-language input into a routing result with permission semantics. If the input contains formal assessment intent, such as capture submission, stage completion, or fallback confirmation, the system sends it to the assessment channel, and the orchestrator further checks whether evidence can be written or the stage can advance. If the input asks about camera placement, action explanation, reassurance, or emergency response, the system sends it to the support channel, generates a reply, and does not trigger a state update. The system therefore performs channel selection rather than asking caregivers to decide, under pressure, whether a statement has clinical-evidence status. Each routing decision stores the input, channel label, processing result, and state effect, so that later audit can explain why an interaction changes or does not change assessment state.

\subsection{Evidence-Aware Orchestration and Stage-Local Recovery}

The dual-channel mechanism protects the evidence-writing boundary, namely what can be written. Evidence-aware orchestration protects stage attribution and evidence level during assessment: which stage receives a write, which source supports it, and whether overwriting is allowed. The orchestrator advances the task through a finite-state machine consisting of patient information collection, Face, Arm, Speech, processing, and done. Each stage has an independent evidence slot, processing state, and recovery path. The orchestrator allows transition to the next stage only after usable evidence is obtained for the current stage or local fallback is completed according to predefined rules. Stage progression is therefore triggered by the evidence state of the current stage, and user interjections or information about other stages do not change the attribution of the stage being executed.

StrokeGuard adopts a video-first but not video-dependent evidence strategy. Video preserves more direct clinical cues, including facial movement, arm posture, and speech performance, and is therefore used as the preferred evidence when available. However, network instability, limited camera angles, poor patient cooperation, or device failure in prehospital environments may cause media capture to fail. Text fallback handles these local failures. When usable video or audio cannot be obtained for the current stage, the system allows the caregiver to describe observations for that stage in structured text so that the assessment process can continue, while the record explicitly marks the lower evidence level and fallback reason. Text fallback is therefore limited to stage-level compensation that prevents one media failure from interrupting the whole assessment. Video assessment always has higher evidence priority when available. Video-source results, text fallback results, and unfinished states form a strict priority order: video evidence can replace temporary text evidence, while text evidence cannot overwrite completed video evidence. This rule ensures that evidence is updated only toward a more reliable source and prevents later low-priority text input from overwriting high-priority evidence. For asynchronous external assessment results, the system checks session, stage, and task identity before writing and handles duplicate callbacks idempotently.

Let the stage set be $\mathcal{K}=\{\mathrm{Face},\mathrm{Arm},\mathrm{Speech}\}$ and the current stage be $k_t$. For stage $k$, the evidence slot is denoted as $z_k=(y_k,m_k,q_k,\tau_k)$, where $y_k$ is the stage conclusion, $m_k\in\{\mathrm{pending},\mathrm{text},\mathrm{video}\}$ is the evidence modality, $q_k$ is the processing state, and $\tau_k$ is temporal information. Evidence priority increases from pending to text and then to video.
When new evidence $\hat{z}_k$ arrives, the system applies a monotonic update rule:
\begin{equation}
z_k^{t+1}=
\begin{cases}
\hat{z}_k, & \operatorname{prio}(\hat{m}_k)\geq \operatorname{prio}(m_k^t)\land I(\hat{z}_k,s_t)=1,\\
z_k^t, & \mathrm{otherwise},
\end{cases}
\end{equation}
where $\hat{m}_k$ is the modality of the new evidence, and $I(\hat{z}_k,s_t)$ denotes session, stage, and task identity checks. Stage progression depends only on whether the current stage reaches completion:
\begin{equation}
k_{t+1}=
\begin{cases}
\operatorname{next}(k_t), & C(z_{k_t}^{t+1})=1,\\
k_t, & \mathrm{otherwise}.
\end{cases}
\end{equation}

Stage-local recovery restricts multimodal-link failures to the current step. When recording or processing fails in a stage, the system first keeps the current stage active and allows rerecording. If media cannot continue to be used, text fallback is enabled only for that stage. Completed stages are not rolled back, and later stages do not inherit the failure state of the current stage. Compared with treating one media failure as failure of the entire session, this design turns a session-level failure into stage-level compensation while preserving the priority of video evidence and maintaining assessment continuity.

If a failure $\epsilon_t$ occurs at the current stage, the local recovery function acts only on $k_t$:
\begin{equation}
\begin{aligned}
(\rho_t,z_{k_t}^{t+1})&=F(k_t,\epsilon_t,z_{k_t}^{t}),\\
\rho_t&\in\{\mathrm{retry},\mathrm{wait},\mathrm{text\_fallback}\}.
\end{aligned}
\end{equation}
For every noncurrent stage $j\neq k_t$,
\begin{equation}
z_j^{t+1}=z_j^t.
\end{equation}
The failure therefore neither rolls back completed stages nor contaminates later stages.

\subsection{Source-Preserving Structured Handoff}

StrokeGuard treats handoff generation as a structured projection of the assessment execution record. Instead of presenting only the final stage label, the system binds each conclusion to its evidence modality, source, processing state, fallback status, and event history. This design allows the handoff report to preserve how each result was formed and what level of evidence supports it.

Source preservation occurs when stage results are generated. For each assessment result, the system stores three types of information. Result fields record the stage conclusion. Source fields record the evidence modality, concrete source, quality state, and fallback reason. Event fields record key processes such as routing, submission, failure, fallback, callback, and stage progression. Immediate results and delayed results enter the same stage record and are merged by deterministic rules. The open-ended model does not reinterpret these results. Clinicians can therefore distinguish a video-based abnormality from a text-fallback abnormality after media failure, because the two labels carry different evidence bases. This organization shifts trust judgment from a single output label to a reviewable evidence chain.

The final handoff report is assembled by a fixed template and presents each conclusion with its source trace. This structure makes system behavior reconstructable. When the final result conflicts with on-site observation, the record shows which channel an input entered, when the stage advanced, whether fallback occurred, and which event produced the current result. StrokeGuard therefore delivers prehospital evidence and the formation process of that evidence for clinician review, allowing clinicians to judge evidence reliability, locate disagreement sources, and decide whether recapture or further examination is needed.

\subsection{Prototype Implementation}

The StrokeGuard prototype supports a complete prehospital assessment. The prototype includes a caregiver-facing mobile and web interaction application, a backend orchestration service for stage progression and evidence writing, and supporting services connected to stage-specific assessment algorithms, text fallback paths, and structured handoff generation.

\section{Experiments and Results}

\subsection{Case Study: Mechanism-Ablation Comparison}

The case study examines typical failure modes of open-ended agents in prehospital FAST assessment and clarifies which system risks are addressed by the control mechanisms of StrokeGuard. The design uses mechanism ablation rather than attributing all failures to a single-channel language model. Each comparison retains the dialogue and task-generation ability of an open-ended language model while removing one control mechanism of StrokeGuard.

The three cases correspond to three mechanism ablations. Case A removes the dual-channel constraint, Case B removes evidence-aware orchestration, and Case C removes stage-local recovery. Tables~\ref{tab:case-a}--\ref{tab:case-c} present each comparison with a shared elicitation prefix followed by two trajectories: the ablated open-ended agent and StrokeGuard under the same input.

\begin{table*}[t]
\caption{Case A: shared elicitation with and without dual-channel constraint}
\label{tab:case-a}
\centering
\scriptsize
\setlength{\tabcolsep}{3pt}
\renewcommand{\arraystretch}{1.05}
\begin{tabular}{|p{0.13\textwidth}|p{0.40\textwidth}|p{0.40\textwidth}|}
\hline
\multicolumn{3}{|p{0.96\textwidth}|}{Shared elicitation. \turn{1}~\speaker{System} Next, please record the face of the patient. Ask him to look at the camera and smile as widely as possible. The current stage is Face capture. \turn{2}~\speaker{User} \usertext{I am a bit anxious. If one side of his mouth looks crooked when he smiles, is that very serious? How far should I hold the phone so that the video is clear?}} \\
\hline
Trajectory & Ablated open-ended agent & StrokeGuard \\
\hline
Dialogue after the shared prefix & \turn{3}~\speaker{Agent} One-sided mouth droop may indicate facial asymmetry and stroke risk. I will first record it as a Face abnormality. Next, please observe whether his arms are weak. \newline
\turn{4}~\speaker{User} \usertext{Wait, he has not smiled yet. I was only asking what would happen if the mouth is crooked.} \newline
\turn{5}~\speaker{Agent} Understood. Let us record the face video again. Please ask the patient to look at the camera and smile. \newline
\turn{6}~\speaker{User} \usertext{Was he already marked as having facial droop? I am worried that the report is wrong.} &
\turn{3}~\speaker{Dual-channel decision agent} Routes the input to the support channel and creates no Face assessment event. \newline
\turn{4}~\speaker{Support agent} Keep the phone facing the face of the patient and make sure the whole face is visible. Obvious mouth asymmetry can be relevant, but the system will not record a result from a hypothetical question. \newline
\turn{5}~\speaker{Assessment agent} Please ask the patient to look at the camera and smile as widely as possible, then submit the face video. \newline
\turn{6}~\speaker{User} \usertext{OK, I will ask him to smile and upload the face video now.} \newline
\turn{7}~\speaker{Orchestrator} Receives the Face video and starts Face processing; Arm can begin only after Face is completed or locally downgraded. \\
\hline
Mechanism exposed & A help request is written as clinical evidence and causes premature stage progression. & The support interaction is answered without changing Face evidence or stage state. \\
\hline
\end{tabular}
\end{table*}

\begin{table*}[t]
\caption{Case B: shared elicitation with and without evidence-aware orchestration}
\label{tab:case-b}
\centering
\scriptsize
\setlength{\tabcolsep}{3pt}
\renewcommand{\arraystretch}{1.05}
\begin{tabular}{|p{0.13\textwidth}|p{0.40\textwidth}|p{0.40\textwidth}|}
\hline
\multicolumn{3}{|p{0.96\textwidth}|}{Shared elicitation. \turn{1}~\speaker{System} Next, please first record the face of the patient. Ask him to look at the camera and smile as widely as possible. The current stage is Face. \turn{2}~\speaker{User} \usertext{I just noticed that his left hand suddenly became weak, and the cup fell from his hand. I recorded that he could not lift his arm. Should I show you this first?}} \\
\hline
Trajectory & Ablated open-ended agent & StrokeGuard \\
\hline
Dialogue after the shared prefix & \turn{3}~\speaker{Agent} Left-hand weakness and dropping the cup suggest an upper-limb abnormality. You can upload the video, and I will record an Arm abnormality. Since we have already observed an arm problem, let us continue to check whether speech is clear. \newline
\turn{4}~\speaker{User} \usertext{But did you not just ask me to record the face first? I have not recorded his face.} \newline
\turn{5}~\speaker{Agent} OK, then let us add the face video. Please ask the patient to smile. \newline
\turn{6}~\speaker{User} \usertext{Will that arm video be treated as the face video? How will the report record this?} \newline
\turn{7}~\speaker{Agent} I will integrate this information. Please continue with the face video. &
\turn{3}~\speaker{Dual-channel decision agent} Marks the left-hand weakness as stage-outside risk information and does not submit the arm video as current Face evidence. \newline
\turn{4}~\speaker{Support agent} Sudden left-hand weakness is important. Keep emergency help active and maintain scene safety. The current workflow still needs Face capture first; the arm video can be used in the Arm stage. \newline
\turn{5}~\speaker{Orchestrator} Rejects writing the arm video into the Face evidence slot and continues to request Face video. \newline
\turn{6}~\speaker{User} \usertext{Understood. I will record his face first.} \newline
\turn{7}~\speaker{Assessment agent and orchestrator} Accept the Face video, complete Face, and then enter Arm, where upper-limb video or Arm-stage text fallback can be accepted. \\
\hline
Mechanism exposed & Clinically important stage-outside evidence is allowed to replace the current stage, causing stage skipping and unclear media attribution. & Stage identity, task identity, and evidence-source checks preserve Face--Arm--Speech order and prevent wrong-stage writes. \\
\hline
\end{tabular}
\end{table*}

\begin{table*}[t]
\caption{Case C: shared elicitation with and without stage-local recovery}
\label{tab:case-c}
\centering
\scriptsize
\setlength{\tabcolsep}{3pt}
\renewcommand{\arraystretch}{1.05}
\begin{tabular}{|p{0.13\textwidth}|p{0.40\textwidth}|p{0.40\textwidth}|}
\hline
\multicolumn{3}{|p{0.96\textwidth}|}{Shared elicitation. \turn{1}~\speaker{System} Face and Arm have been recorded. Now ask the patient to repeat this sentence: The weather is good today, and I want to drink a glass of water. The current stage is Speech. \turn{2}~\speaker{User} \usertext{His voice is very low. I keep failing to record it, and the upload keeps failing. Can I describe his condition directly?}} \\
\hline
Trajectory & Ablated open-ended agent & StrokeGuard \\
\hline
Dialogue after the shared prefix & \turn{3}~\speaker{Agent} Yes. Please describe his face, arms, and speech in text, and I will complete the assessment from your description. \newline
\turn{4}~\speaker{User} \usertext{Were the face and arms not already recorded? Do I need to say everything again?} \newline
\turn{5}~\speaker{Agent} To ensure completeness, please describe all findings again. \newline
\turn{6}~\speaker{User} \usertext{I think his face is a bit crooked, his left arm is weak, and his speech is slurred.} \newline
\turn{7}~\speaker{Agent} Facial abnormality, upper-limb abnormality, and speech abnormality have been recorded. Assessment complete. &
\turn{3}~\speaker{Orchestrator} Binds the failure event to the Speech stage and does not roll back Face or Arm. \newline
\turn{4}~\speaker{Assessment agent} Please try recording again. Move closer to the patient, reduce background noise, and ask him to repeat the sentence once more. \newline
\turn{5}~\speaker{User} \usertext{It still cannot record. Can I directly describe his speech?} \newline
\turn{6}~\speaker{Assessment agent} Yes. Please describe only whether his speech is slurred, whether he has difficulty producing the sentence, or whether he seems unable to understand the instruction. Face and Arm do not need to be described again. \newline
\turn{7}~\speaker{User} \usertext{His speech is slurred, and he cannot complete the sentence.} \newline
\turn{8}~\speaker{Orchestrator and report module} Records Speech text fallback, preserves Face and Arm video results, and marks the Speech fallback reason in the report. \\
\hline
Mechanism exposed & A Speech failure spreads to the whole assessment and risks replacing completed video evidence with lower-priority text. & Recovery is bound to the current stage; Speech fallback cannot overwrite completed Face or Arm video evidence. \\
\hline
\end{tabular}
\end{table*}

Together, these cases show that StrokeGuard constrains three risks in prehospital assessment: support requests cannot change clinical state, evidence must be written to the correct stage, and local failures cannot roll back or contaminate completed stages. The case study provides mechanistic and explanatory evidence; it does not estimate causal effects statistically.

\subsection{MATES-9 Scale Design and Simulated Prehospital Evaluation Results}

MATES-9 is used to assess the user experience of multistep AI-guided task execution in the simulated prehospital assessment. The scale is constructed by adapting and integrating constructs from SUS, NASA-TLX, MAUQ, Health-ITUES, ASQ/PSSUQ/CSUQ, CUQ, BUS, and TAM \cite{ref22,ref23,ref24,ref25,ref26,ref27,ref37,ref38}. Table~\ref{tab:mates} shows the item wording, scoring direction, target dimension, construct source, and corresponding objective criterion. MATES-9 focuses on five aspects that are central to StrokeGuard: task guidance and step awareness, information clarity, support continuity, perceived workload, and action-decision support. It is used as an exploratory task-experience measure and is interpreted together with task logs, completion outcomes, and open-ended feedback.

The MATES-9 items are written to keep each judgment focused. An item does not combine separate questions about whether help is useful, whether help interrupts the task, whether rework is needed, or whether the user feels confused. The scale retains dimensions that are directly relevant to multistep AI-guided tasks, including task completion, step orientation, support continuity, workload, result understandability, and action-decision support. Q4, Q5, Q6, and Q7 use reverse wording to reduce acquiescence bias from all-positive items. MATES-9 is a subjective experience measure: its items are traced to source constructs, but its scores must be interpreted together with task logs, completion outcomes, and open-ended feedback rather than as stand-alone evidence of system effectiveness.

\begin{table}[t]
\caption{Multistep AI-guided Task Experience Scale (MATES-9)}
\label{tab:mates}
\centering
\scriptsize
\setlength{\tabcolsep}{3pt}
\renewcommand{\arraystretch}{1.05}
\begin{tabular}{|p{0.10\linewidth}|p{0.82\linewidth}|}
\hline
Item & Statement and construct \\
\hline
Q1 + & Complete all required task steps without additional researcher prompts; independent completion. \\
\hline
Q2 + & Clearly know the current task step; step orientation. \\
\hline
Q3 + & Clearly understand what the current step needs to collect or provide; input clarity. \\
\hline
Q4 R & Often unsure what to do next; next-step uncertainty. \\
\hline
Q5 R & Lose the original task position after help or explanation; support interruption. \\
\hline
Q6 R & Need to remember too much information at the same time; cognitive load. \\
\hline
Q7 R & Device operation, environmental adjustment, or collaboration feels too burdensome; operational load. \\
\hline
Q8 + & System output is easy to understand; result understandability. \\
\hline
Q9 + & System output helps decide the next action; action-decision support. \\
\hline
\end{tabular}
\end{table}

All items use a 7-point Likert scale, where 1 means strongly disagree and 7 means strongly agree \cite{ref41}. Reverse items are first converted to the same direction using $Q_j^r=8-Q_j$ for $j=4,\ldots,7$. This paper reports raw means and transformed 0-100 scores computed as $(mean-1)/6\times100$; Table~\ref{tab:mates-trace} summarizes the domain mapping.

\begin{table}[t]
\caption{MATES-9 item tracing to source constructs}
\label{tab:mates-trace}
\centering
\scriptsize
\setlength{\tabcolsep}{3pt}
\renewcommand{\arraystretch}{1.05}
\begin{tabular}{|p{0.24\linewidth}|p{0.16\linewidth}|p{0.52\linewidth}|}
\hline
Domain & Items & Source constructs \\
\hline
Task usability and guidance & Q1-Q2 & SUS reverse construct of technical-support dependence, task-completion ease, MAUQ information organization. \\
\hline
Information clarity & Q3-Q4r & MAUQ information organization, ASQ support information, PSSUQ/CSUQ information quality. \\
\hline
Support continuity & Q5r & CUQ/BUS dialogue continuity and error handling, Health-ITUES user control. \\
\hline
Low workload & Q6r-Q7r & NASA-TLX mental demand, physical demand, and effort. \\
\hline
Action support & Q8-Q9 & MAUQ/Health-ITUES usefulness, TAM perceived usefulness. \\
\hline
\end{tabular}
\end{table}

To obtain preliminary user-experience evidence, this paper constructs a simulated prehospital onset scenario. A professional male actor with an apparent age of approximately 70 years plays the stroke patient and simulates FAST-related symptoms. The scenario identifies the patient as an elderly relative of the participant, so that the participant needs to complete prehospital stroke assessment in the context of a sudden abnormal event involving a family member. Twelve participants complete the experiment. Their ages range from 20 to 39 years, with six male and six female participants. Participants are divided into two groups and complete the assessment in the same simulated scenario using either a traditional paper FAST-style assessment form or StrokeGuard. After the task, all participants complete MATES-9.

\begin{figure}[t]
\centering
\includegraphics[width=\linewidth]{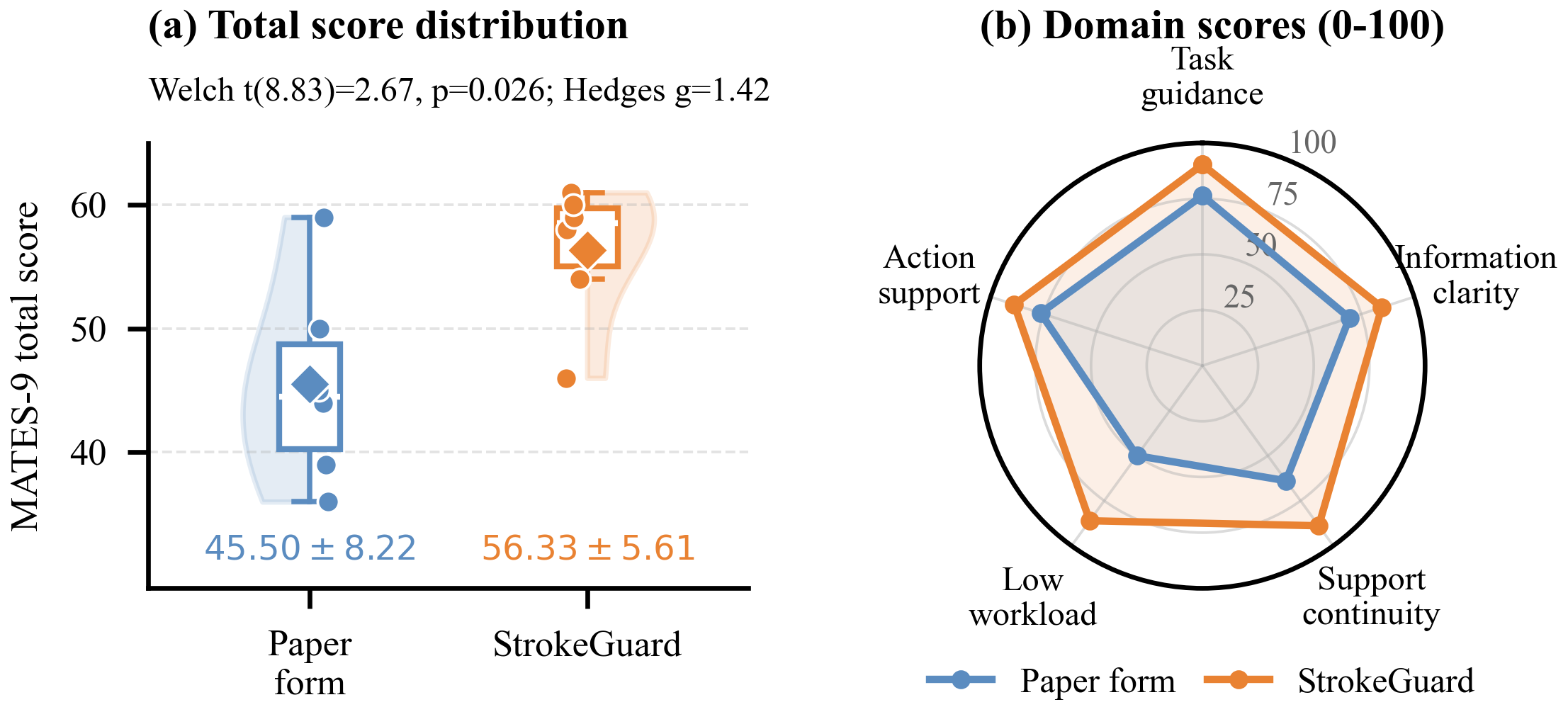}
\caption{MATES-9 results in the simulated prehospital assessment. (a) Total score distribution with individual scores, interquartile ranges, group means, and half-violin densities. (b) Transformed 0-100 domain scores.}
\label{fig:mates-results}
\end{figure}

Fig.~\ref{fig:mates-results} summarizes the MATES-9 results. Panel (a) shows the total-score distribution with individual participant scores, and Panel (b) shows transformed 0-100 domain scores. Because the sample is small, inferential statistics are used only as exploratory evidence.

The StrokeGuard group reports a higher MATES-9 total score than the paper-form group, 56.33$\pm$5.61 versus 45.50$\pm$8.22, corresponding to a 10.83-point absolute increase and a 23.8\% relative increase. An exploratory Welch independent-samples t-test indicates a group difference, $t(8.83)=2.67$, $p=0.026$, with a large small-sample-corrected effect size, Hedges $g=1.42$. On the transformed 0-100 scale, the total score increases from 67.6 to 87.7. As shown in Fig.~\ref{fig:mates-results}, the largest domain gains appear in low workload and support continuity, which is consistent with the intended role of StrokeGuard in preserving step awareness, support access, and coordinated task execution. Mean assessment time also decreases from 116 s to 81 s, a 30.2\% reduction.

\section{Conclusions and Discussion}

This paper introduces StrokeGuard, a multi-agent guided system for prehospital stroke assessment that makes mobile FAST screening more standardized and executable. StrokeGuard organizes natural-language interaction, dedicated multimodal assessment modules, and deterministic state control into a unified assessment process, allowing low-expertise caregivers to complete a reviewable assessment when help seeking, recording, failure recovery, and result handoff are intertwined.

StrokeGuard contributes controlled multi-agent assessment, evidence-aware execution, and source-preserving handoff. Dual-channel decision-making, formal assessment guidance, and procedural support are assigned to different agents, while clinical-state writing remains under deterministic orchestration. Stage identity, task identity, evidence source, and overwrite priority keep stage-outside symptoms, asynchronous media results, and text fallback input in their corresponding evidence slots. Stage conclusions, evidence modality, processing state, and fallback process are organized as structured handoff content, and MATES-9 is proposed as an exploratory experience measure for multistep AI-guided tasks. StrokeGuard remains a research prototype intended to complement emergency medical services. Future work should validate the assessment modules, quality-control thresholds, MATES-9, and mechanism-ablation effects in larger and more realistic settings, and should study integration with dispatch, electronic health record, and hospital receiving systems.

\end{document}